\documentclass[%
 reprint,
superscriptaddress,
 amsmath,amssymb,
 aps,
pra,
]{revtex4-2}

\usepackage[ruled,vlined]{algorithm2e}
\usepackage{url}
\usepackage{amsfonts,amsmath, amsthm, amscd,amssymb,amsbsy,amstext,amsopn,amsxtra}
\usepackage{hyperref}
\usepackage{float}  
\usepackage{subcaption}
\usepackage[justification=raggedright,singlelinecheck=false]{caption}
\usepackage{tikz}
\usepackage{siunitx}

\usepackage{fancyhdr}
\usepackage{graphicx}
\usepackage{dcolumn}
\usepackage{bm}

\begin{document}

\preprint{APS/123-QED}

\title{A scalable superconducting circuit framework for emulating physics in hyperbolic space}

\author{Xicheng Xu}
\thanks{These authors contributed equally (Joint first authors)}
\affiliation{Institute for Quantum Computing (IQC), University of Waterloo, Waterloo, Canada}
\affiliation{Department of Physics and Astronomy, University of Waterloo, Waterloo, Canada}
\affiliation{Red Blue Quantum Inc., Waterloo, Canada}

\author{Ahmed Adel Mahmoud}
\thanks{These authors contributed equally (Joint first authors)}
\affiliation{Centre for Quantum Topology and Its Applications (quanTA), University of Saskatchewan, Saskatoon, Canada}
\affiliation{Department of Mathematics and Statistics, University of Saskatchewan, Saskatoon, Canada}

\author{Noah Gorgichuk}
\affiliation{Institute for Quantum Computing (IQC), University of Waterloo, Waterloo, Canada}
\affiliation{Department of Physics and Astronomy, University of Waterloo, Waterloo, Canada}
\affiliation{Red Blue Quantum Inc., Waterloo, Canada}

\author{Ronny Thomale}
\affiliation{Institut für Theoretische Physik und Astrophysik, Universität Würzburg, Würzburg, Germany}

\author{Steven Rayan}
\thanks{These authors supervised the project (Joint supervising authors).}
\affiliation{Centre for Quantum Topology and Its Applications (quanTA), University of Saskatchewan, Saskatoon, Canada}
\affiliation{Department of Mathematics and Statistics, University of Saskatchewan, Saskatoon, Canada}
\affiliation{Qaleidoscope Intelligence Inc., Toronto, Canada}

\author{Matteo Mariantoni}
\thanks{These authors supervised the project (Joint supervising authors).}
\affiliation{Institute for Quantum Computing (IQC), University of Waterloo, Waterloo, Canada}
\affiliation{Department of Physics and Astronomy, University of Waterloo, Waterloo, Canada}
\affiliation{Red Blue Quantum Inc., Waterloo, Canada}

\begin{abstract}
Theoretical studies and experiments in the last six years have revealed the potential for novel behaviours and functionalities in device physics through the synthetic engineering of negatively-curved spaces. For instance, recent developments in hyperbolic band theory have unveiled the emergence of higher-dimensional eigenstates—features fundamentally absent in conventional Euclidean systems~\cite{maciejko2021hyperbolic, maciejko2022automorphic, kienzle2022hyperbolic}. At the same time, superconducting quantum circuits have emerged as a leading platform for quantum analogue emulations and digital simulations in scalable architectures~\cite{boettcher2020quantum, schmidt2013circuit, zhang2023superconducting}. Here, we introduce a scalable superconducting circuit framework for the  analogue quantum emulation of tight-binding models on hyperbolic and kagome-like lattices. Using this approach, we experimentally realize three distinct lattices, including—for the first time to our knowledge—a hyperbolic lattice whose unit cell resides on a genus-3 Riemann surface. Our method encodes the hyperbolic metric directly into capacitive couplings between high-quality superconducting resonators, enabling tenable reproduction of spectral and localization properties while overcoming major scalability and spectral resolution limitations of previous designs. These results set the stage for large-scale experimental studies of hyperbolic materials in condensed matter physics and lay the groundwork for realizing hyperbolic quantum processors, with potential implications for both fundamental physics and quantum computing. 
\end{abstract}

\maketitle


\section{Introduction}\label{main}
The study of physical phenomena in hyperbolic spaces, once primarily the domain of high-energy physics and cosmology~\cite{chen2003hyperbolic,kehagias2000hyperbolic}, has recently emerged as a vibrant frontier in condensed matter and quantum computing. This paradigm shift has been catalysed by a series of theoretical breakthroughs, including recent advances in hyperbolic band theory~\cite{maciejko2021hyperbolic, maciejko2022automorphic}, crystallography of hyperbolic lattices~\cite{boettcher2022crystallography}, hyperbolic topological insulators~\cite{wu2015topological, liu2023higher, liu2022chern} and hyperbolic quantum error correction codes~\cite{albuquerque2009topological, breuckmann2016constructions, higgott2024constructions}. These advances have spurred the experimental efforts in emulating hyperbolic materials, opening the door to investigating fundamentally new phenomena that are inaccessible in conventional Euclidean geometries~\cite{kollar2019hyperbolic, lenggenhager2022simulating, huang2024hyperbolic, chen2023hyperbolic, zhang2022observation, zhang2023hyperbolic, patino2024hyperbolic, dey2024simulating, bienias2022circuit, park2024scalable}. These experiments have used different platforms to perform analogue emulations of hyperbolic materials. Notably, one of the earliest efforts was based on superconducting coplanar waveguide (CPW) resonators to emulate the spectrum of a tight-binding model on a hyperbolic lattice~\cite{kollar2019hyperbolic}. Despite its conceptual novelty, this approach faced three important limitations: (1) The emulations were restricted to effective lattices with kagome-like connectivity rather than actual hyperbolic tilings. (2) The spectral resolution of their experimental data was insufficient to enable qualitative comparison with the theoretical spectrum. (3) It relied on emulating the inhomogeneity of the hyperbolic metric by varying the physical distances between resonators, significantly hindering scalability. In particular, the lattice spacing shrinks exponentially with radial distance from the centre, when embedded into flat two-dimensional space. This exponential contraction severely limits the number of realizable lattice layers, as components become too densely packed beyond a few concentric shells.   

Subsequent experiments, using room-temperature topoelectric circuits, aimed to emulate the spectral properties as well as topological states of certain hyperbolic lattices~\cite{lenggenhager2022simulating, zhang2022observation, zhang2023hyperbolic}. While innovative, these approaches faced fundamental limitations stemming from the inhomogeneity of the hyperbolic metric, and the low quality factor (Q-factor) of the room-temperature, normal-conducting resonators precluded fine resolution of the spectral features. 

In this work, we introduce a new framework for emulating hyperbolic lattices that overcomes the key limitations of previous approaches. Our method encodes the inhomogeneity of the hyperbolic metric into the coupling capacitances between high Q-factor superconducting resonators. This approach enables us to: 

\begin{enumerate}
\def\labelenumi{\arabic{enumi}.}
\item
Realize genuine hyperbolic lattices, contrary to the earlier assumption that CPW resonator architectures could only emulate kagome-like lattices~\cite{kollar2019hyperbolic}.
\item
  Reproduce the energy spectrum of hyperbolic tight-binding models with better resolution. 
\item
  Establish a scalable framework that supports significantly larger lattices, as it is not constrained by the exponential contraction of inter-site physical distances inherent to hyperbolic embeddings. 
\end{enumerate}

One of the central contributions of this work is the experimental distinction between kagome-like lattices and their parent hyperbolic lattices. This distinction is revealed by the presence of a highly degenerate ground state (flat band) in the measured energy spectrum of the kagome-like lattice—a feature absent in its hyperbolic counterpart. In addition, we present, to the best of our knowledge, the first emulation of a hyperbolic lattice whose unit cell is embedded in a genus-3 Riemann surface. Realizing such lattices experimentally is particularly challenging because the tiling polygons have a large number of edges. This leads to two main difficulties: first, the dense spatial arrangement required to accommodate many resonators within each polygon, and second, the increased vertex connectivity, which demands precise control over a greater number of coupling elements. While current device packaging constraints limit our study to relatively modest system sizes, the underlying framework is intrinsically scalable to larger implementations. 

\section{The Tight-Binding Model on Hyperbolic and Kagome-like Lattices}\label{the-tight-binding-model}

Interest in hyperbolic lattices resurged with recent advances in hyperbolic band theory. For periodic Euclidean lattices, Bloch’s theorem guarantees that the eigenstates of the tight-binding Hamiltonian can be expressed in terms of momentum states. This allows obtaining exact solutions via methods grounded in group theory and the theory of unitary representations of translational symmetry. However, recent advances in hyperbolic band theory have revealed a richer structure for the eigenstates of the tight-binding Hamiltonian defined on hyperbolic lattices~\cite{maciejko2021hyperbolic, maciejko2022automorphic}. Firstly, the momentum space is no longer two-dimensional but a higher-dimensional space. In this case, while some eigenstates correspond to one-dimensional irreducible representations and thus adhere to Bloch’s theorem, others correspond to higher-dimensional representations that fall outside this scope. These higher-dimensional representations emerge from the non-Abelian nature of the Fuchsian group that defines the lattice symmetries~\cite{katok1992fuchsian}, fundamentally distinguishing hyperbolic systems from their Euclidean counterparts~\cite{nagy2024hyperbolic}. 

A related class of interest is kagome-like lattices, which have gained increasing attention in recent years due to their unique geometric and spectral properties~\cite{kollar2020line, bzdusek2022flat}. These lattices are constructed from parent hyperbolic or Euclidean lattices by placing a vertex at the midpoint of every edge and connecting vertices if their parent edges share a common vertex. This procedure enhances vertex connectivity; for example, kagome-like lattices derived from regular $\{p,3\}$ lattices have vertices with uniform degree-4 connectivity, in contrast to the degree-3 connectivity of their parent lattices. We discuss the spectral properties of these lattices in more details in the Methodology~\ref{sec:kagome-and-flatband}. 

On the other hand, the Hamiltonian operator plays a central role in describing the dynamics of non-relativistic quantum systems. For interacting quantum systems, the Hamiltonian is generally a non-linear operator whose exact solution is often intractable. To address this challenge, a common approach involves replacing the full interacting Hamiltonian with a corresponding non-interacting one, which often captures a significant portion of the system’s dynamics~\cite{ashcroft1976solid}. In the non-interacting case, the Hamiltonian is typically expressed as the sum of two terms: a kinetic term, represented by the Laplacian operator $\Delta$, and a potential term $V$, so that $H = -\Delta + V$. In the context of crystalline solids, two standard limits are considered: (1) The free electron model, where the kinetic energy dominates and the Hamiltonian reduces to $H = -\Delta.$ (2) The tight-binding model, where the binding energy dominates, yielding $H=V$~\cite{kittel2004introduction}.

The tight-binding approximation is particularly effective for capturing the low-energy part of the system’s spectrum. When the lattice sites are sufficiently well-separated relative to the spatial width of the electron bound states, the solutions are typically expressed as linear combinations of the single-electron bound states. These bound states are localized at each atomic core and propagate through weak quantum tunnelling between these cores. This leads to a transition from a continuum model of the crystal to a discrete lattice model that captures the spectrum of the continuum~\cite{boettcher2020quantum}. The corresponding tight-binding Hamiltonian takes the form: 
\begin{equation}
\label{eqn:hamiltonian}
H = - t \sum_{i,j} A_{ij} a_i^\dagger a_j,
\end{equation}
where \(A_{ij}\) is the adjacency matrix that encodes the connectivity between vertices and thereby capture the hyperbolic structure of the lattice, while \(a_{i}^{\dagger}\) and \(a_{j}\) are the creation and
annihilation operators at lattice sites $i$ and $j$ respectively. This formulation provides a natural starting point for studying electronic spectra of hyperbolic lattices.

\begin{figure}[h!]
    \centering    
    \includegraphics[]{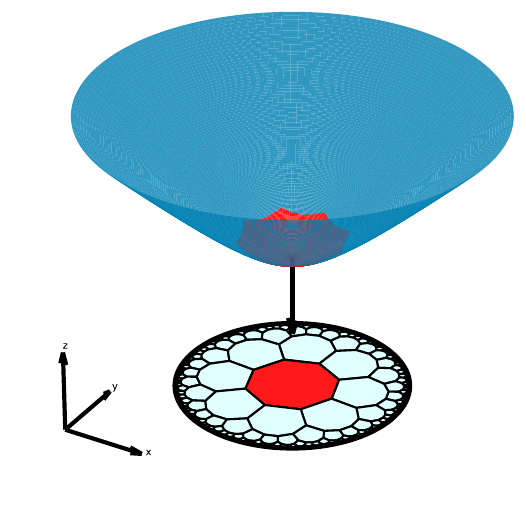}
    \caption{Stereographic projection from the hyperbolic space to the
Poincaré disk. The disk is tiled by the regular $\{8,3\}$ hyperbolic lattice, with the reference octagon (red) in the hyperbolic space projected onto the unit cell of the tiling.}
    \label{fig:fig1_projection}
\end{figure}

In this work, we conduct three experiments to probe the spectral properties of the tight-binding Hamiltonian defined on hyperbolic and kagome-like lattices. We employ the Poincaré disk model of the hyperbolic space, in which the lattice is embedded within a unit disk equipped with a non-uniform hyperbolic metric. We focus on regular hyperbolic $\{p,q\}$ lattices, which are tessellations of the Poincaré disk by regular $p$-gons, with $q$ $p$-gons meeting at each vertex. Figure.~\ref{fig:fig1_projection} shows the stereographic projection from the hyperbolic space to the Poincaré disk tiled by a regular $\{8,3\}$ hyperbolic lattice. 

\section{Experiments}
\label{sec:experiments}

We experimentally realize hyperbolic lattices on two-dimensional devices using a superconducting circuit architecture based on semi-lumped-element CPW resonators operated in the microwave range. The tight packing of the meandering lines adds significant capacitance, making it semi-lumped-element. By cooling them to a base temperature of approximately \SI{10}{mK}, the resonators are initialized in the quantized ground (vacuum) state. In this state, they can be represented by bosonic modes~\cite{schmidt2013circuit,carusotto2020photonic,zhang2023superconducting},  making them naturally suited for emulating tight-binding Hamiltonians, with their interactions described by  Eq.~\ref{eqn:hamiltonian}. Once the high-power measurement begins, the system behaves classically, and its response can be described with classical equations. Nevertheless, the system remains consistent with the quantized Eq.~\ref{eqn:hamiltonian}, as the eigenvalues of Eq.~\ref{eqn:hamiltonian} can still predict the resonance peak frequencies, which is computationally simpler than solving equations of coupled classical resonators. Furthermore, because our framework is inherently compatible with circuit quantum electrodynamics (circuit QED) architectures, Eq.~\ref{eqn:hamiltonian} can be readily extended to include qubits or other fully quantum mechanical elements. In this case, specifically, we choose superconducting resonators to serve as the lattice vertices (atomic sites), and capacitive couplings to mimic the edges (interatomic distances), with the coupling strengths encoding the physical distances between sites. This choice relies on the fact that the packing of the meandering semi-lumped-element CPW resonators makes their footprint much smaller than the wavelength of its fundamental mode. Therefore, it can be treated as a point-like object, making it an excellent analogue for an atomic site. In particular, we chose CPW resonators for their high Q-factors, which provide far greater spectral resolution than normal-metal or even lumped element superconducting resonators~\cite{bejanin2022fluctuation}. Furthermore, half-wave CPW resonators are employed for their flexibility to be coupled at both ends of the waveguide (see Extended Data Fig.~\ref{fig:figS1_couplers}), whereas full-wave CPW resonators have larger footprints and quarter-wave CPW resonators are more difficult to couple to because one end is shorted to ground.

As described earlier, our framework encodes the hyperbolic metric into the capacitive couplings between resonators. We achieve this by engineering the coupling capacitances to be inversely proportional to the physical distances between the lattice vertices represented by the resonators. Two coupling strategies are possible: Direct coupling, where resonators are capacitively linked end-to-end, and mediated coupling, where small capacitive islands (couplers) are placed between resonator ends. In our experiments, the hyperbolic lattices employ direct capacitive coupling, whereas couplers are used to construct the kagome-like lattices. Compared with the coupler-based method, direct coupling is inherently more scalable, as neighbouring resonators can support a wide range of ratios of coupling capacitances. Consequently, the coupling strengths within a single device can span several orders of magnitude.

The spectral properties of the lattice manifest as peaks in the transmission data. For reference, when measuring a single half-wave CPW resonator with both ends coupled to the ports, the transmission spectrum exhibits a peak when the probing frequency matches its energy eigenvalue. The same principle applies to the entire lattice, where a peak in the transmission spectrum indicates a resonance corresponding to an energy eigenvalue of the system. In particular, the $n$th theoretical eigenvalue, $\lambda_n$, is mapped to its corresponding frequency $f_n$ such that
\begin{equation}
\label{eqn:frequencies}
    f_n = f_1 + (\lambda_n-\lambda_1)\frac{f_2 - f_1}{\lambda_2 - \lambda_1}, \quad n \geq 1,
\end{equation}
where $f_1$ and $f_2$ are the first two frequencies associated with the first two theoretical eigenvalues. Note that $f_1$ and $f_2$ are based on the locations of the first two peaks in the experimental data, and $\lambda_n \ (n \geq 1)$ are the theoretical eigenvalues of Eq.~\ref{eqn:hamiltonian}. While the ideal tight-binding spectrum of the lattice contains degenerate eigenstates that should correspond to the same frequency $f_n$, minor fabrication imperfections are expected to lift these degeneracies in the measured samples. This effect manifests as clusters of closely spaced peaks, where highly degenerate states translate to higher number of peaks in close proximity. In the measured data, to keep the analysis consistent, for peaks to feature a cluster they have to exhibit three features: (1) There  exists more than one peak in the cluster. (2) At least one of these peaks has a high transmission value (usually $>$ \SI{-40}{dB}). (3) They are well-separated from other clusters of peaks.

To experimentally probe the spectral properties of the lattice, each device is fabricated with four interfaces: three serving as the input ports and one as the output port for transmission measurements, enabling us to measure the scattering parameters, $S_{21}$, $S_{31}$, and $S_{41}$. The use of multiple ports increases the likelihood of exciting and detecting a larger subset of eigenstates, as the spatial profile of certain eigenstates may prevent them from being illuminated by certain input ports.  Consequently, the experimentally observed spectrum contains only those eigenstates with sufficient spatial coupling to at least one of the input–output channels. In addition, device imperfections such as crosstalk between the resonators can lead to spurious couplings, producing extra peaks and shifting the frequencies of existing peaks, thereby distorting the spectral features. To recover as much spectral information as possible, we aggregate the data from all three input ports ($S_{21}$, $S_{31}$, and $S_{41}$) by taking the maximum transmission value at each frequency point. In the subsections that follow, we present the results of applying this methodology to three representative lattices: the $\{8,3\}$ lattice, the $\{12,4\}$ lattice, and the kagome-like $\{8,3\}$ lattice. The data for the intermediate devices are presented in Extended Data Figs.~\ref{fig:figS5_12gon_kagome}, \ref{fig:figS6_12gon_Eu}, and \ref{fig:figS7_8gon_Eu}.

\subsection{The $\{8,3\}$ Lattice Experiment}
\begin{figure*}
    \centering
    \includegraphics[width=1\linewidth]{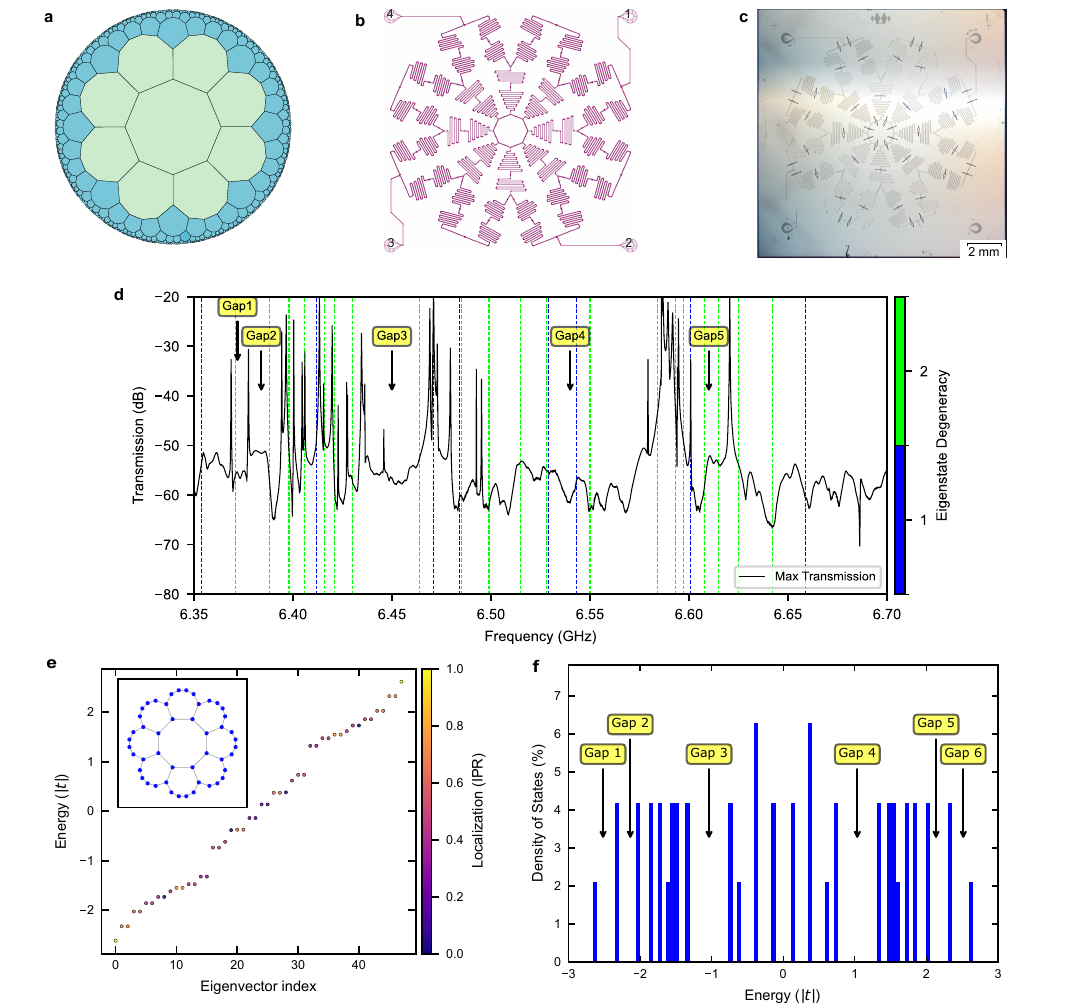}
    \caption{\textbf{a}, The $\{8,3\}$ lattice in the Poincaré disk model, with the emulated sublattice highlighted in light green. \textbf{b}, Sample design of the $\{8,3\}$ sublattice displayed in (a), composed of 48 semi-lumped element half-wave CPW resonators ($\sim$\SI{6.5}{GHz} fundamental resonance frequency)  with direct capacitive couplings. \textbf{c}, The micrograph of the measured device (1.5cm$\times$1.5cm) with visible wire bounds. \textbf{d}, Measured transmission spectrum, obtained by taking the maximum of $S_{21}$, $S_{31}$, and $S_{41}$ at each frequency. Vertical lines indicate the mapped theoretical eigenvalues, colour-coded by degeneracy, with the maximum degeneracy being 2. Five gaps are identified between clusters of peaks. \textbf{e}, The tight binding spectrum of the sublattice in (a) obtained via numerical diagonalization. The colourmap encodes the inverse participation ratio (IPR), which quantifies the degree of localization of each eigenstate. \textbf{f}, Density of states (DOS),  histogram for the spectrum in (e) computed with a bin width of $0.03|t|$. The DOS is symmetric about $E=0$, with vertical bars showing the normalized DOS at each discrete eigenvalue; the annotated arrows mark spectral gaps (Gap 1–Gap 6), including two large gaps (Gap 3, Gap 4) separating the three main clusters of states and narrower gaps (Gap 1, Gap 2, Gap 5, and Gap 6) near the band edges. Only gaps with widths exceeding $0.25|t|$ are highlighted.}
    \label{fig:fig2_8gon_actual}
\end{figure*}

Our first experiment focuses on a sample designed to emulate the regular $\{8,3\}$ hyperbolic lattice---a tiling of the Poincaré disk by octagons, with three octagons meeting at each vertex. The unit cell of this lattice can be compactified on a genus-2 Riemann surface~\cite{boettcher2022crystallography}. Our sample realizes this structure with 9 faces, 48 vertices, and 56 edges, as shown in the micrograph of Fig.~\ref{fig:fig2_8gon_actual}b. The corresponding sample features 48 resonators, each one directly coupled to its nearest neighbours in the lattice. The tight-binding spectrum, obtained by numerical diagonalization, is presented in Fig.~\ref{fig:fig2_8gon_actual}e, while its density of states (DOS) is shown in Fig.~\ref{fig:fig2_8gon_actual}f. The DOS is symmetric around the energy $E= 0$ across the spectrum with two large gaps around the energies \(E =  \pm 1\) and two smaller gaps near each end.

The experimental data are shown in Fig.~\ref{fig:fig2_8gon_actual}d, where all 48 energy eigenvalues from diagonalization are also overlaid with colour-coded degeneracy. The procedure to map the theoretical eigenvalues to the spectrum follows Eq.\ref{eqn:frequencies}. The three individual S-parameter data and the spectrum in a wider window are presented in Extended Data Fig.~\ref{fig:figS2_8gon_actual}. We now interpret the data both qualitatively and quantitatively.

Qualitatively, the experimental results in Fig.~\ref{fig:fig2_8gon_actual}d align well with the theoretical spectrum represented by the DOS in Fig.~\ref{fig:fig2_8gon_actual}f. The DOS shows three large clusters of eigenstates separated by two prominent gaps (Gap 3 and Gap 4). These clusters correspond to the three groups observed in the measured data, centred around the frequencies 6.41, 6.47, and 6.59 GHz, with the gaps appearing near 6.45 and 6.54 GHz. In addition, the smaller gaps (Gap 1, Gap 2, and Gap 5) in Fig.~\ref{fig:fig2_8gon_actual}f also have counterparts in the measured data near 6.37, 6.38, and 6.61 GHz in Fig.~\ref{fig:fig2_8gon_actual}d, while Gap 6 is absent in the experimental data.  

Quantitatively, the first two-fold degenerate eigenstates near 6.37 GHz align well with the corresponding theoretical predictions, indicated by the overlaid green lines. Similarly, the peaks in the cluster around 6.40 GHz match multiple two-fold degenerate eigenstates. The next peaks, centred at 6.47 GHz, also agrees with the overlaid theoretical spectrum, consisting of three two-fold and one non-degenerate eigenstate. In contrast, the eigenstates predicted between 6.50 and 6.57 GHz are not observed experimentally. The two-fold degenerate eigenstates at higher frequencies align with the cluster near 6.60 GHz, while the final two eigenstates--—one of which is two-fold degenerate—--are likewise missing from the measured data.

\subsection{The $\{12,4\}$ Lattice Experiment}
\begin{figure*}
    \centering
    \includegraphics[width=1\linewidth]{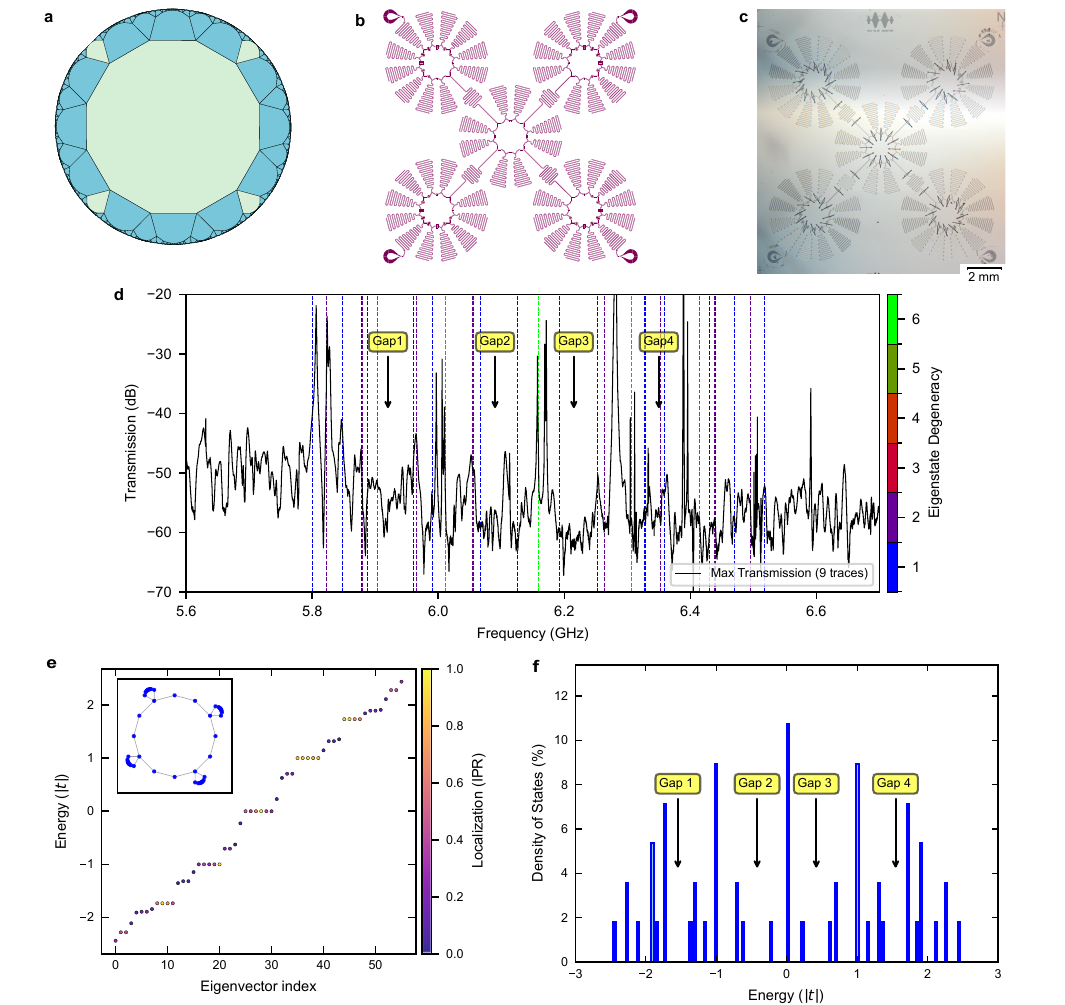}
     \caption{\textbf{a}, The $\{12,4\}$ lattice in the Poincaré disk model, with the emulated sublattice highlighted in light green. The unit cell of this lattice corresponds to a genus-3 Riemann surface. \textbf{b}, Sample design of the $\{12,4\}$ lattice, composed of 56 semi-lumped element half-wave CPW resonators ($\sim$\SI{6.5}{GHz} fundamental resonance frequency) with direct capacitive couplings. \textbf{c}, The micrograph of the measured device (1.5cm$\times$1.5cm) with visible wire bounds. \textbf{d}, Measured transmission spectrum, obtained by taking the maximum of the 9 sets of data of $S_{21}$, $S_{31}$, and $S_{41}$, across 3 different cooldowns, at each frequency. Vertical lines indicate the mapped theoretical eigenvalues, colour-coded by degeneracy, with the highest degeneracy being 6. There is good alignment between highly degenerate eigenstates and clusters of peaks, while low degenerate eigenstates are harder to identify in the measured data. Four gaps are identified between clusters of peaks. \textbf{e}, The tight binding spectrum of the sublattice in (a), obtained via numerical diagonalization. The colourmap encodes IPR, which quantifies the degree of localization of each eigenstate. \textbf{f}, Density of states histogram for the spectrum in (e) computed with a bin width of $0.03|t|$. The DOS is symmetric about $E=0$, with vertical bars showing the normalized DOS at each discrete eigenvalue. The graph features five clusters separated by four mid-size gaps.  Only gaps with widths exceeding $0.2|t|$ are highlighted.}
    \label{fig:fig3_12gon_actual}
\end{figure*}

The second experiment aims at emulating a hyperbolic lattice whose unit cell corresponds to a genus-3 Riemann surface~\cite{boettcher2022crystallography}. The goal of this experiment is to show to versatility of this experimental framework and explore a new territory of lattices that has not been explored before. For that purpose, we emulated a regular $\{12,4\}$ hyperbolic lattice; this lattice features dodecagons, where every 4 dodecagons meet at each vertex. The sample features 5 dodecagons, 56 vertices and 60 edges. Correspondingly, the chip features 56 resonators, each coupled directly to its nearest neighbours. As before, the tight-binding spectrum obtained via diagonalization is shown in Fig.~\ref{fig:fig3_12gon_actual}e., while the DOS is shown in Fig.~\ref{fig:fig3_12gon_actual}f. The three individual S-parameter data and the spectrum in a wider window are presented in Extended Data Fig.~\ref{fig:figS3_12gon_actual}. Similar to the $\{8,3\}$ sample, we analyze the data qualitatively and quantitatively.

Qualitatively, the measured spectrum in Fig.\ref{fig:fig3_12gon_actual}d agrees well with the theoretical DOS in Fig.\ref{fig:fig3_12gon_actual}f. The DOS is symmetric about $E=0$and exhibits five clusters separated by four gaps. Based on the criteria given for clusters of peaks in Experiments~\ref{sec:experiments}, the measured data displays five clusters of peaks within the 5.8–-6.4~GHz range, reflecting the same clustered structure and gaps as in Fig.\ref{fig:fig3_12gon_actual}f.  
 
Quantitatively, the theoretical spectrum in Fig.~\ref{fig:fig3_12gon_actual}e exhibits five highly degenerate eigenstates: one six-fold degenerate state at $E = 0$, two five-fold degenerate states at $E = \pm 1$, and two four-fold degenerate states at $E = \pm 1.73$. These states are reflected in the experimental data in Fig.~\ref{fig:fig3_12gon_actual}d, whereas eigenstates with lower degeneracy are mostly absent. Specifically, the eigenstate at $E = 0$ corresponds to the cluster at 6.18 GHz; the two eigenstates at $E = \pm 1$ correspond to the clusters at 6.01 and 6.31 GHz; and the eigenstates at $E = \pm 1.73$ map onto the clusters at 5.82 and 6.39 GHz. Overall, the agreement between theory and experiment is reasonable, with the exception of the state at $E =-1.73$, which appears shifted a higher frequency. This deviation can be attributed to the experimental imperfections discussed earlier.

\subsection{The Kagome-like $\{8,3\}$ Lattice Experiment}
\begin{figure*}
    \centering
    \includegraphics[width=1.0\linewidth]{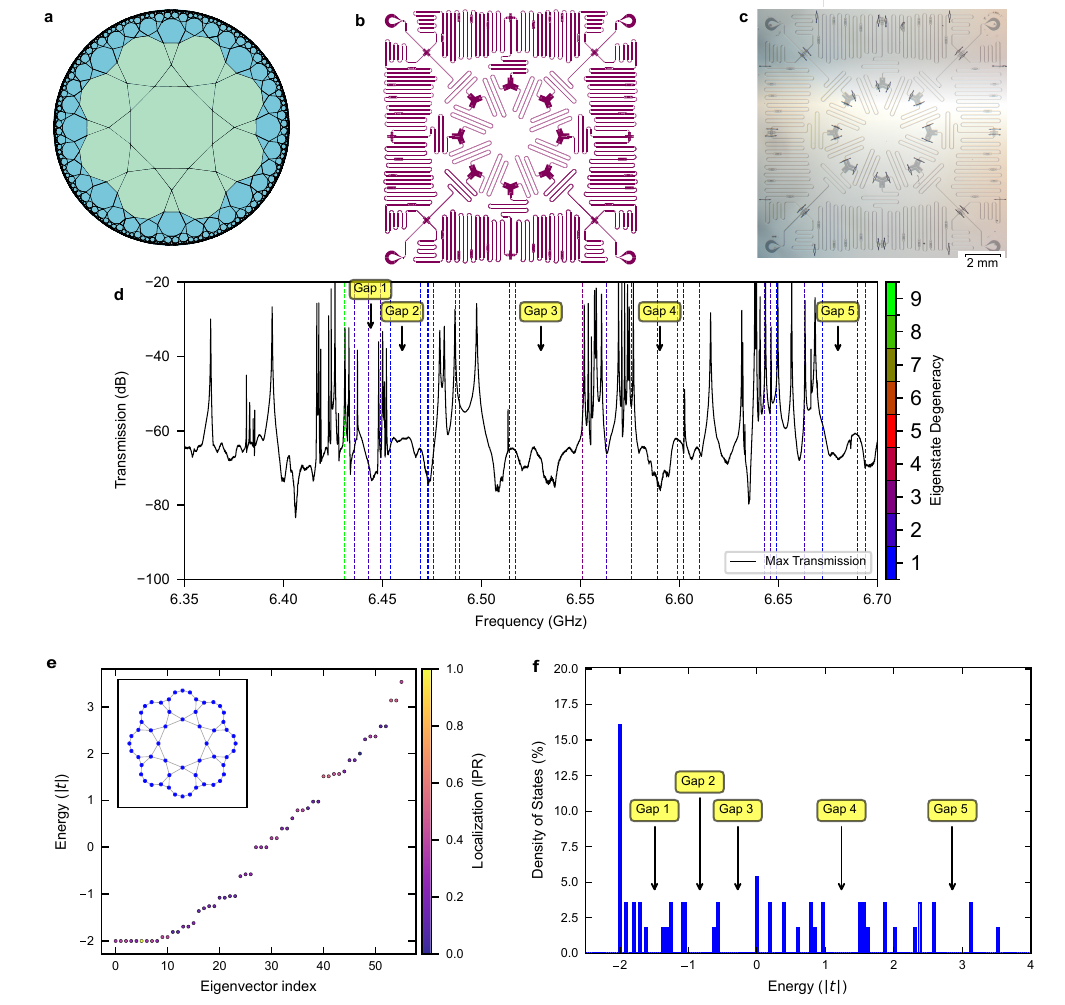}
     \caption{\textbf{a}, The kagome-like lattice, inherited from an $\{8,3\}$ lattice, in the Poincaré disk model, with the emulated sublattice highlighted in light green. \textbf{b}, Sample design of the $\{8,3\}$ kagome-like lattice, composed of 56 semi-lumped element half-wave CPW resonators ($\sim$\SI{6.5}{GHz} fundamental resonance frequency) coupled through capacitive couplers. The design features uniform degree-4 connectivity between resonators. \textbf{c}, The micrograph of the measured device (1.5cm$\times$1.5cm) with visible wire bounds. \textbf{d}, Measured transmission spectrum, obtained by taking the maximum of $S_{21}$, $S_{31}$, and $S_{41}$ at each frequency. Vertical lines indicate the mapped theoretical eigenvalues, colour-coded by degeneracy, with the highest degeneracy being 9 at the ground state. Five gaps are identified between clusters of peaks. \textbf{e}, The tight binding spectrum of the sublattice in (a), obtained via numerical diagonalization. It features a 9-fold degenerate ground state representing the flat band. The colormap encodes the IPR, which quantifies the degree of localization of each eigenstate. \textbf{f}, Density of states histogram for the spectrum in (e) computed with a bin width of $0.03|t|$. The vertical bars showing the normalized DOS at each discrete eigenvalue. As expected from a kagome-like lattice, the DOS is no longer symmetric about $E=0$ as in the previous samples. Instead, the flat band constitutes a large part of the spectrum (around 16\%).}
    \label{fig:fig4_8gon_kagome}
\end{figure*}

The third experiment emulates an $\{8,3\}$ kagome-like lattice that features 56 vertices and 80 edges. This lattice is constructed from a parent $\{8,3\}$ lattice by placing a vertex on each edge of the parent lattice and connecting two vertices if their underlying edges share a common vertex. The device design is shown in Fig.~\ref{fig:fig4_8gon_kagome}b; the sample consists of 56 resonators, coupled via 48 2-way, 3-way or 4-way couplers. The three individual S-parameter data and the spectrum in a wider window are presented in Extended Data Fig.~\ref{fig:figS4_8gon_kagome}. As before, we interpret the data both qualitatively and quantitatively.

Qualitatively, the DOS in Fig.~\ref{fig:fig4_8gon_kagome}f exhibits six distinct clusters: three wide and three narrow. The wide clusters appear just before Gap 1 and just after Gaps 3 and 4, while the narrow clusters occur just after Gaps 1, 2, and 5. The measured data in Fig.~\ref{fig:fig4_8gon_kagome}d shows good overall agreement with this structure. Specifically, the three wide clusters manifest as broad peak clusters around the frequencies 6.43, 6.57, and 6.65 GHz, while two of the narrow clusters appear as sharper features near 6.45 and 6.48 GHz. While the absence of the final narrow cluster after Gap 6 may be due to the imperfections stated above, it is also possible that eigenvalues near the highest eigenvalues are more difficult to resolve experimentally. The peaks below 6.4 GHz are not classified as a cluster because their low transmission amplitudes in accordance with the criteria given in the Experiments~\ref{sec:experiments}.

\begin{figure}[t]
    \centering
    \includegraphics[width=1.0\linewidth]{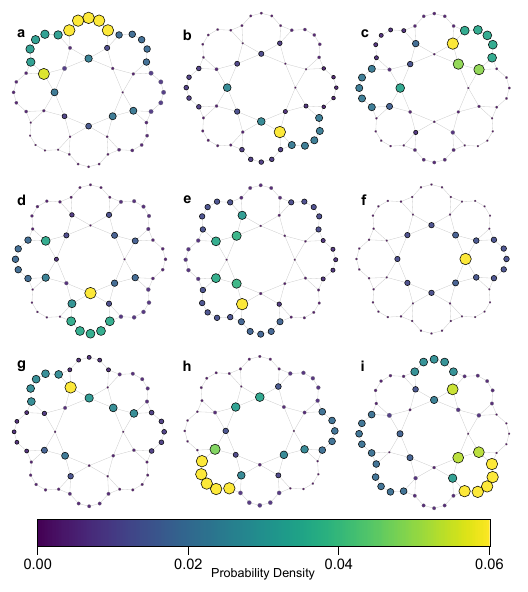}
    \caption{Compact localized states associated with the flat-band ground state of the kagome-like lattice shown in Fig.~\ref{fig:fig4_8gon_kagome}a. Each CLS exhibits spatial localization with finite support on a small subset of lattice sites, with eigenstate \textbf{f} displaying the highest degree of localization. Node size and color encode the probability density of the corresponding eigenstate, as indicated by the colour bar. Nodes outlined with black borders mark vertices, where the probability density exceeds 0.01.}
    \label{fig:fig5_CLSs}
\end{figure}

Quantitatively, the parent $\{8,3\}$ lattice is flat (i.e., it does not employ periodic boundary conditions) and contains nine plaquettes, each hosting a maximally localized compact localized state (CLS). Consequently, the kagome-like lattice is expected to feature a 9-fold degenerate ground state. The 9-fold degeneracy of the ground state is confirmed in the theoretical spectrum obtained via diagonalization (Fig.~\ref{fig:fig4_8gon_kagome}e), while the eigenstates are illustrated in Fig.\ref{fig:fig5_CLSs}. This flat band is well captured in the measured data through the cluster consisting of exactly nine peaks centred at the frequency 6.43 GHz. Moreover, 2-fold and 3-fold degenerate eigenstates are reasonably well aligned with the clusters present at higher frequencies likes the ones near 6.49 and 6.64 GHz. In contrast, some of the non-degenerate eigenvalues do not align well with the transmission peaks as the one near 6.59 GHz, due to similar imperfection mechanisms to the other two devices.    

\section{Conclusion}\label{conclusion}
In this work, we introduced a new framework for emulating hyperbolic lattices using superconducting resonators operated in the microwave range. The novelty of our approach lies in encoding the inhomogeneity of the hyperbolic metric directly into the capacitive couplings between resonators, unlike previous approaches, which mimicked hyperbolic geometry by varying the physical distances between resonators. Using this method, we emulated three lattices: two hyperbolic and one kagome-like. Among these is the first example of a hyperbolic lattice whose unit cell is embedded on a genus-3 Riemann surface. Our framework also overcomes several limitations of earlier emulations. It enables the realization of genuine hyperbolic lattices, alongside kagome-like geometries, and faithfully reproduces features of their tight-binding spectra. Finally, our framework allows for higher scalability due to the wide tunability of capacitances across multiple orders of magnitude, compared to the dense packing of resonators hindering further scalability of previous approaches.  

Even though our method is inherently scalable, our current packaging limitations restricted the samples to mid-size lattices. Further studies that exploit this approach to emulate larger lattices are still needed. Another challenge is imposing periodic boundary conditions on the emulated hyperbolic lattices. This approach would enable the exploration of exotic phenomena in hyperbolic lattices predicted by hyperbolic band theory, including the emergence of higher-dimensional eigenstates of the tight-binding Hamiltonian. An additional avenue for future research lies in the connection between hyperbolic lattices and holographic principles~\cite{boyle2020conformal, asaduzzaman2020holography, asaduzzaman2022holography}. Unlike Euclidean lattices, where the ratio of boundary to bulk nodes vanishes in the thermodynamic limit, hyperbolic lattices preserve a finite boundary-to-bulk ratio. This unique geometric property provides a natural setting for probing holographic phenomena within a controlled experimental platform. Realizing this goal will require access to larger lattices, for which our framework provides a viable experimental foundation. Finally, since our framework is compatible with circuit QED architectures, another direction is to include superconducting qubits into the circuit with the goal to explore the error correcting capabilities of hyperbolic quantum error correction codes. Even though these codes have been of increasing interest on the theoretical level, experimental validity of these codes is still missing. 
\newpage

\section{Methodology}
\label{sec:method}

\subsection{Sample Design}
\label{sec:sample-design}

Because the hyperbolic plane is stereographically projected onto the Poincaré disk (Fig.~\ref{fig:fig1_projection}), the spacing between neighbouring vertices becomes non-uniform: it is largest at the central polygon and exponentially decreases toward the boundary. While the absolute values of these distances are not fixed, their ratios are strictly determined by the geometry of the lattice. This property introduces flexibility: The overall scale can be freely chosen; meanwhile, the relative proportions remain preserved. To exploit this, we designate a single reference coupling capacitance that is compatible with the fabrication limits and the device size. Since coupling capacitances are inversely proportional to the distances between the vertices, the chosen reference corresponds to the largest distance, namely that of the central polygon. This ensures that the smallest capacitance is set first, after which the remaining capacitances can be systematically derived by applying the predetermined ratios. In this way, the geometric constraints of the hyperbolic tiling are naturally translated into a consistent set of device parameters.

Because performing the finite element simulation of the entire circuit consisting of semi-lumped element CPW resonators is too computationally demanding, we use SPICE, a lumped element circuit simulator, to simulate the lumped element circuit model consisting of capacitors and inductors~\cite{pozar2021microwave} to estimate the S-parameters results (Extended Fig.~\ref{fig:figS2_8gon_actual}, \ref{fig:figS3_12gon_actual}, \ref{fig:figS4_8gon_kagome}, \ref{fig:figS5_12gon_kagome}, \ref{fig:figS6_12gon_Eu},and \ref{fig:figS7_8gon_Eu}). We next employ an iterative procedure to determine the precise geometry of the resonators and their coupling points. This process consists of repeated electromagnetic simulations using \emph{Ansys Q3D Extractor}, a quasi-static field solver designed to compute capacitances. At each step, the geometry of the resonators is adjusted, and the resulting capacitances are compared against the target values set by the distance ratios. The procedure continues until the simulated capacitances converge to the desired specifications. 

In addition to achieving the required coupling strengths, the device layout is simultaneously optimized for compactness, ensuring that the entire circuit can be accommodated within the \SI{1.5}{cm}$\times$\SI{1.5}{cm} device area, limited by the packaging. Special attention is paid to maintaining sufficient clearances between non-neighbouring resonators, to suppress spurious couplings and minimize crosstalk. The outcome of this process is a layout that balances capacitance precision with physical feasibility, providing a scalable and experimentally viable implementation of the hyperbolic lattices.

\subsection{Fabrication and Measurement}
\label{sec:fabrication-and-measurement}

All devices were fabricated using the same process. Bare silicon wafers were first cleaned with RCA-1 and hydrogen fluoride~\cite{earnest2018substrate}, followed by the deposition of a 100-nm aluminium film using an electron-beam evaporator (MEB 550 SL3-UHV, Plassys). The photolithography step was performed with S1811 photoresist and a maskless aligner (MLA 150, Heidelberg). The aluminium was then etched with Type A aluminium etchant, with etching parameters adjusted in each run to account for variations in native oxide growth. After resist removal, extensive aluminium wire bonds were applied to interconnect the ground planes, ensuring a uniform potential and suppressing spurious electromagnetic modes. Although the wire bonds introduce small inductances, the circuit elements are insensitive to magnetic fields, and the dense bonding provides high redundancy, as shown in the micrographs in Fig.~\ref{fig:fig2_8gon_actual}b, Fig.~\ref{fig:fig3_12gon_actual}b, and Fig.~\ref{fig:fig4_8gon_kagome}b. The completed devices were then mounted in a custom sample package~\cite{bejanin2016three}.

We performed all measurements as follows. The devices were cooled below 10 mK in a dilution refrigerator. Each of the three input ports was connected to an individual RF line with $\sim$\SI{70}{dB} total attenuation, while one port was connected to the readout chain with $\sim$\SI{70}{dB} total gain. S-parameters were then recorded using a vector network analyzer (N5242A, Keysight). To ensure consistency, the probe power from the analyzer was fixed at \SI{0}{dBm}; tests at other powers produced no significant changes apart from variations in noise levels. To capture all the features (i.e., peaks), from the S-parameters, we processed the transmission data by taking, at each frequency point, the maximum amplitude among $S_{21}$, $S_{31}$, and $S_{41}$. Since certain eigenstates couple weakly or not at all to specific ports, this procedure increases the likelihood of detecting the eigenstates. 

\subsection{Kagome-like Lattices and Flat Bands}
\label{sec:kagome-and-flatband}
Kagome-like lattices have gained significant attention in recent years due to their unique geometric and spectral properties. These lattices can be systematically derived from both Euclidean and hyperbolic tilings by placing a vertex at the centre of each edge of the parent lattice and connecting the vertices whose parent edges share a common vertex. This construction increases the vertex connectivity. A key feature of a kagome-like lattice is the presence pf highly degenerate ground state, whose energy is independent of the crystal momentum, giving rise to a dispersionless flat band~\cite{kollar2020line, bzdusek2022flat}. In the tight-binding description, the energy of this flat band takes the value $E = -2t$, where $t$ denotes the hopping amplitude in Eq.~\ref{eqn:hamiltonian}. For convenience, setting $t=1$, a kagome-like lattice derived from a $\{p,3\}$ parent tiling yields a spectrum spanning the interval $[-2,4)$, with the flat band located precisely at $E=-2$ Fig.~\ref{fig:fig4_8gon_kagome}e. This should be contrasted with the parent $\{p,3\}$ lattice, whose spectral range is $(-3, 3)$. 

The physical significance of the flat band lies in the nature of its eigenstates. These eigenstates are highly localized, with wavefunctions confined to only a few lattice sites, and are therefore referred to as compact localized states. These states not only govern the low-energy physics of the system but also reflect the underlying graph structure of the lattice. For a hyperbolic lattice, the cycle structure of the associated graph plays a central role. There are two types of topological cycles: trivial cycles, which correspond to plaquettes or products of plaquettes, and non-trivial cycles, which cannot be expressed as such products. For a lattice with $F$ plaquettes, there exist $F-1$ linearly independent trivial cycles, each associated with a plaquette. In addition, there are $2g$ linearly independent non-trivial cycles, where $g$ is the genus of the underlying Riemann surface. Each linearly independent cycle corresponds to a distinct CLS. More precisely, there exists $F-1$ CLSs, each having a non-vanishing support on a unique plaquette and $2g$ CLSs, each having a non-vanishing support on a non-trivial cycle, up to equivalence. This establishes a direct link between the spectral properties of the CLSs and the graph-theoretic cycle space of the lattice. The basis of this cycle space consists of all plaquette cycles (excluding one) together with the $2g$ non-trivial cycles, such that any other cycle can be expressed as a linear combination of these elements~\cite{mahmoud2025systematic}. This construction highlights the fundamental role played by CLSs in both spectral and combinatorial descriptions of kagome-like lattices.  

In the absence of periodic boundary conditions, the kagome-like lattice lacks nontrivial cycles, and the plaquette cycles cease to be linearly dependent. Consequently, each face of the parent lattice contributes one linearly independent cycle. Therefore, the number of CLSs for a kagome-like lattice, derived from a parent flat lattice, equals the number of plaquettes in the parent lattice. The emulated kagome-like lattice in Fig.~\ref{fig:fig4_8gon_kagome}a is inherited from a parent $\{8,3\}$ lattice, which is flat and possesses nine plaquettes. Hence, the associated kagome-like lattice is expected to exhibit a nine-fold degenerate ground state. This prediction can be directly confirmed by the spectrum in Fig.~\ref{fig:fig4_8gon_kagome}e computed via exact diagonalization. Another reason for the significance of the flat band is that its eigenstates account for a significant portion of the total spectrum. For the emulated lattice, they constitute approximately $16\%$ of all eigenstates, as illustrated in Fig.~\ref{fig:fig4_8gon_kagome}f. The support of these nine eigenstates is depicted in Fig.~\ref{fig:fig5_CLSs}. In contrast, the spectrum of the parent hyperbolic lattice is markedly less structured and does not exhibit a highly degenerate ground state. This lack of high spectral degeneracy makes the corresponding features more challenging to resolve experimentally. 

\section*{Acknowledgement}
We would like to thank Jan Kycia and Joseph Maciejko for helpful discussions. We also would like to thank the developers of the \texttt{hypertiling} package~\cite{schrauth2023hypertiling}, which was used to generate the lattice figures presented in this work.

X.X., N.G., and M.M. acknowledge funding from the Canada First Research Excellence Fund (CFREF) and the support of the Natural Sciences and Engineering Research Council of Canada (NSERC). A.A.M. and S.R. acknowledge support from the NSERC Discovery Grant program and the Canada Foundation for Innovation (CFI) John R. Evans Leaders Fund (both held by S.R.).

\bibliography{bib}

\newpage
\appendix
\section{Extended Data Figures}
\begin{figure}[h!]
    \centering    
    \includegraphics{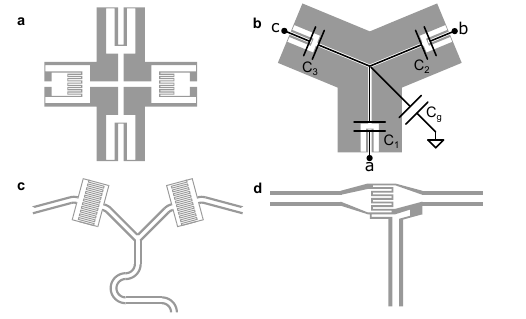}
    \caption{Different example coupling schemes. \textbf{a}, A 4-way coupler. \textbf{b}, A 3-way coupler with the circuit model overlaid. The ratio of couplings is calculated as $g_{ab}/g_{ac} = \frac{C_1(C_2+C_g)}{C_1+C_2+C_g} / \frac{C_1(C_2+C_g)}{C_1+C_2+C_g}$, whose maximum is 2. \textbf{c} and \textbf{d}, Two examples of the direct coupling scheme.}
    \label{fig:figS1_couplers}
\end{figure}

\begin{figure*}[h!]
    \centering    
    \includegraphics{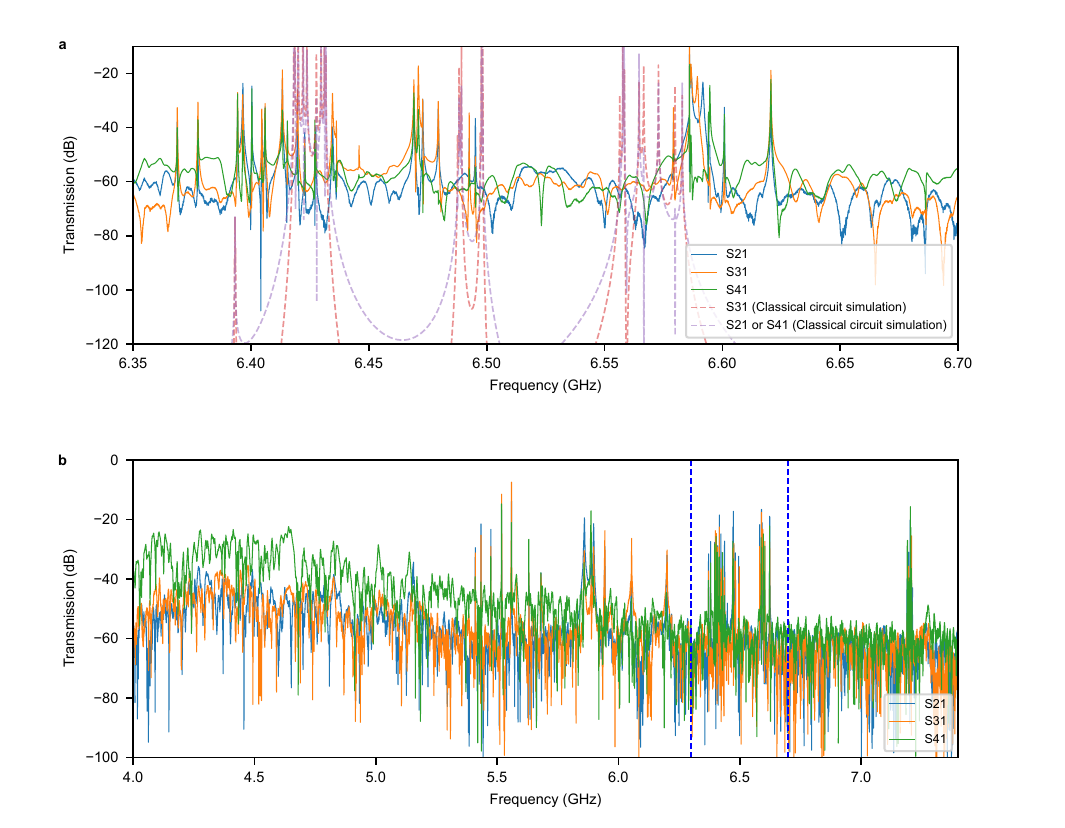}
    \caption{Additional data of the $\{8,3\}$ lattice. \textbf{a}, The unprocessed S-parameters, and the results from the classical circuit simulations with LTSpice. Resonators are modelled as LC resonators coupled via capacitors. \textbf{b}, The same S-parameters in the widest spectrum window allowed by the cryogenic measurement setup, where the dotted blue lines show the region in (a) and Fig.2d.This plot illustrates that there is no more dense area of features apart from what is analyzed in the $\{8,3\}$ lattice subsection in Sec.\ref{sec:experiments}.}
    \label{fig:figS2_8gon_actual}
\end{figure*}

\begin{figure*}[h!]
    \centering    
    \includegraphics{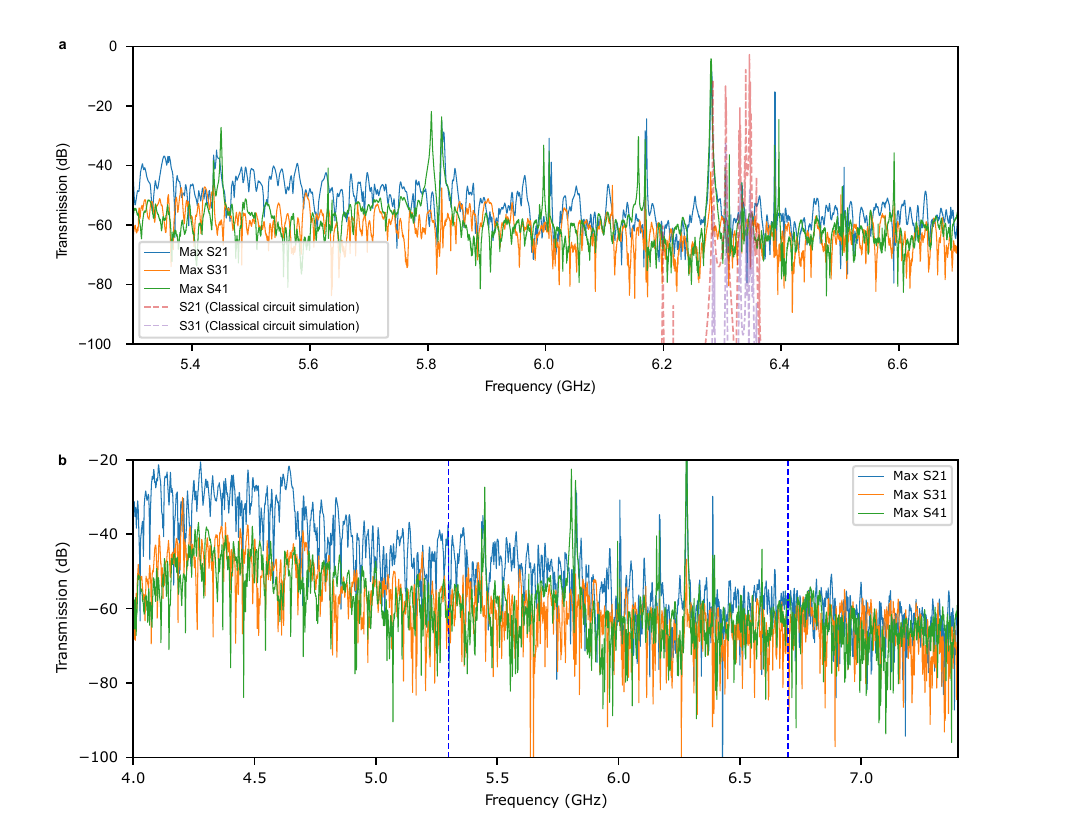}
    \caption{Additional data of the $\{12,4\}$ lattice. \textbf{a}, Transmission spectra obtained by taking, at each frequency, the maximum among the measured S-parameters for three different cool-downs, along with results from classical circuit simulations using SPICE. The resonators are modeled as LC oscillators coupled via capacitors.  \textbf{b}, The same S-parameters in the widest spectrum window allowed by the cryogenic measurement setup, where the dotted blue lines show the region in (a) and Fig.\ref{fig:fig3_12gon_actual}d. This plot illustrates that there is no more dense area of features apart from what is analyzed in the $\{12,4\}$ lattice subsection in  Sec.\ref{sec:experiments}.}
    \label{fig:figS3_12gon_actual}
\end{figure*}

\begin{figure*}[h!]
    \centering    
    \includegraphics{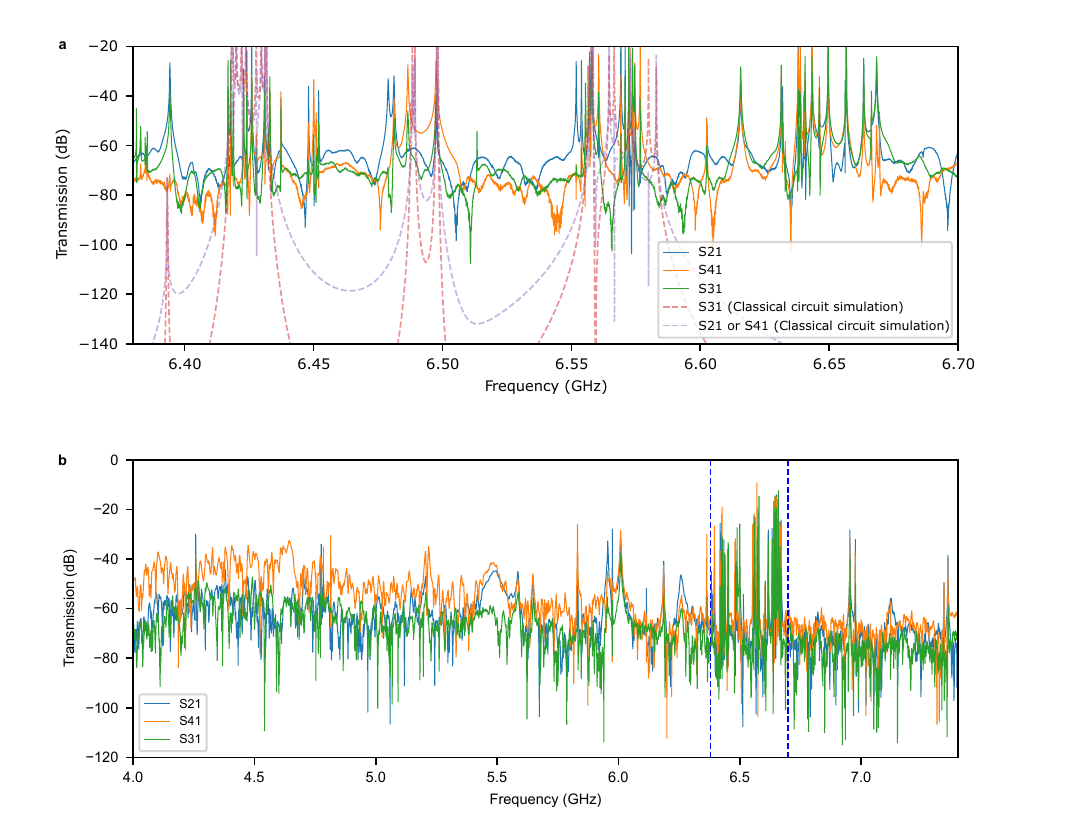}
    \caption{Additional data of the kagome-like $\{8,3\}$. \textbf{a}, The unprocessed S-parameters, and the results from the classical circuit simulations with SPICE. Resonators are modeled as LC resonators coupled via capacitors. \textbf{b}, the same S-parameters in the widest spectrum window allowed by the cryogenic measurement setup, where the dotted blue lines show the region in (a) and Fig.\ref{fig:fig4_8gon_kagome}d. This plot illustrates that there is no more dense area of features apart from what is analyzed in the kagome-like $\{8,3\}$ lattice subsection in Sec.\ref{sec:experiments}.}
    \label{fig:figS4_8gon_kagome}
\end{figure*}

\begin{figure*}[h!]
    \centering    
    \includegraphics{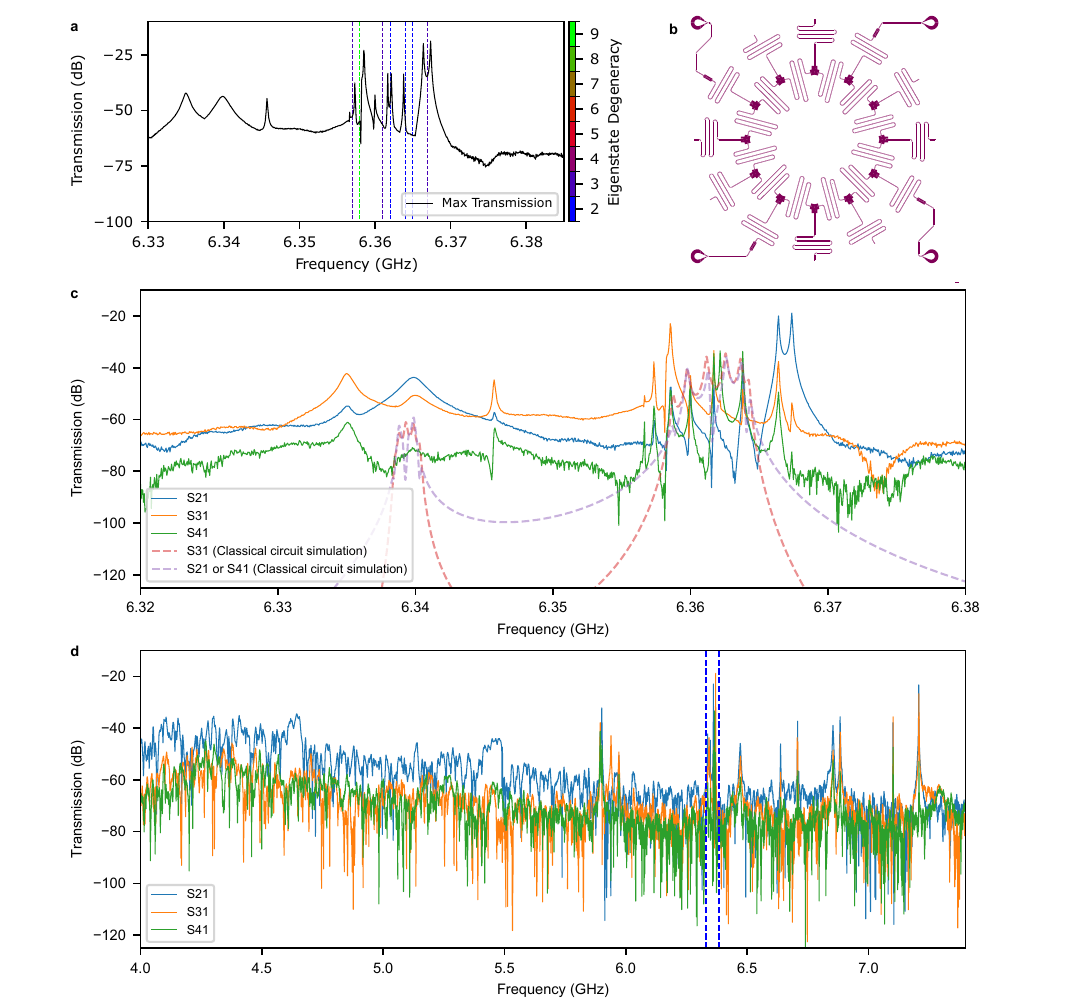}
    \caption{Data and the sample design of a kagome-like $\{12,4\}$ lattice. \textbf{a}, Measured transmission spectrum, obtained by taking the maximum of $S_{21}$, $S_{31}$, and $S_{41}$ at each frequency. The eigenvalues are mapped with the same method described in Sec.\ref{sec:experiments}. \textbf{b}, The sample design. \textbf{c}, The unprocessed S-parameters, and the results from the classical circuit simulations with SPICE. Resonators are modeled as LC resonators coupled via capacitors. \textbf{d}, The same S-parameters in the widest spectrum window allowed by the cryogenic measurement setup, where the dotted blue lines show the region in (c). This plot illustrates that there is no more dense area of features apart from what is analyzed in (c).}
    \label{fig:figS5_12gon_kagome}
\end{figure*}

\begin{figure*}[h!]
    \centering    
    \includegraphics{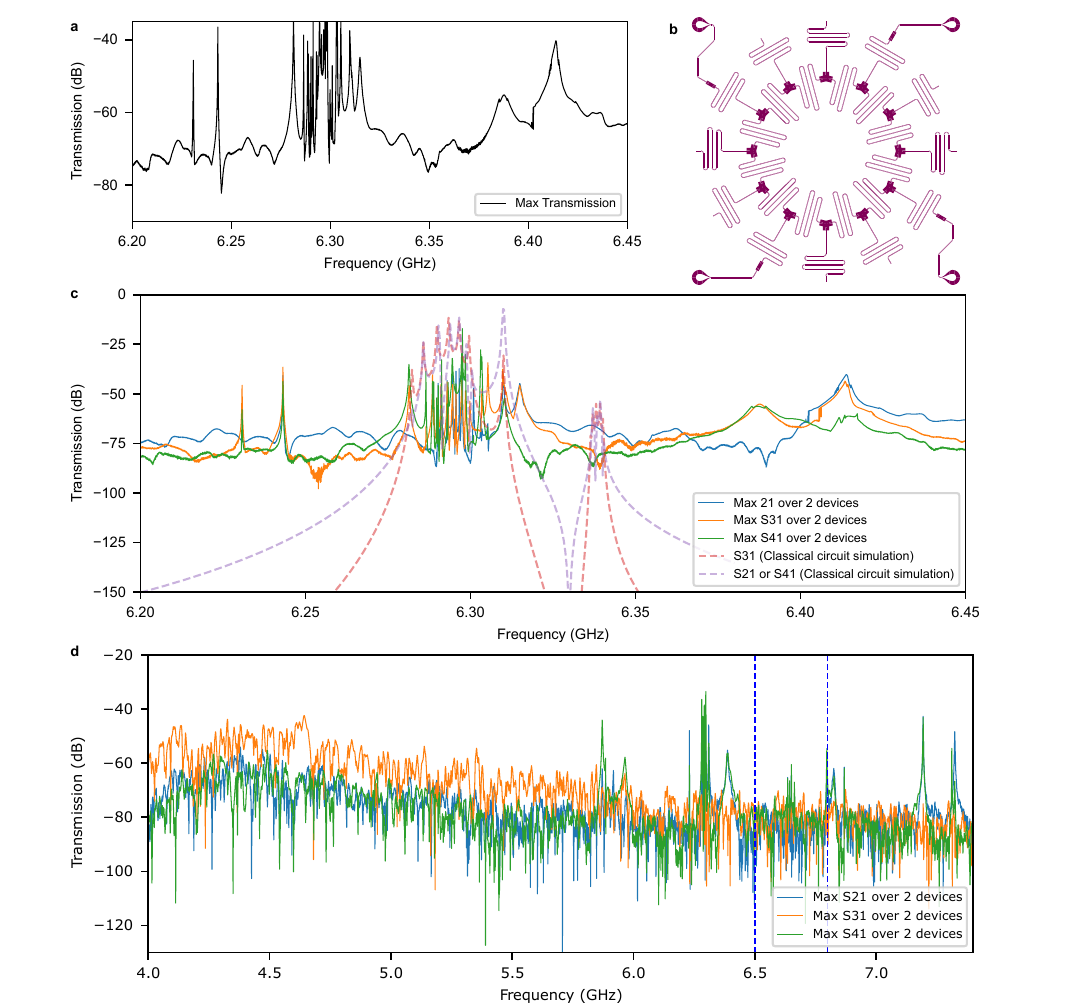}
    \caption{Data and the sample design of a kagome-like $\{12,4\}$ lattice with equal coupling between all neighbouring sites, which we call Euclidean coupling. \textbf{a}, Transmission spectra obtained by taking, at each frequency, the maximum among the measured S-parameters for $S_{21}$, $S_{31}$, and $S_{41}$. \textbf{b}, The sample design. \textbf{c}, The unprocessed S-parameters, and the results from the classical circuit simulations with SPICE. Resonators are modelled as LC resonators coupled via capacitors. \textbf{d}, The same S-parameters in the widest spectrum window allowed by the cryogenic measurement setup, where the dotted blue lines show the region in (c). This plot illustrates that there is no more dense area of features apart from what is analyzed in (c).}
    \label{fig:figS6_12gon_Eu}
\end{figure*}

\begin{figure*}[h!]
    \centering    
    \includegraphics{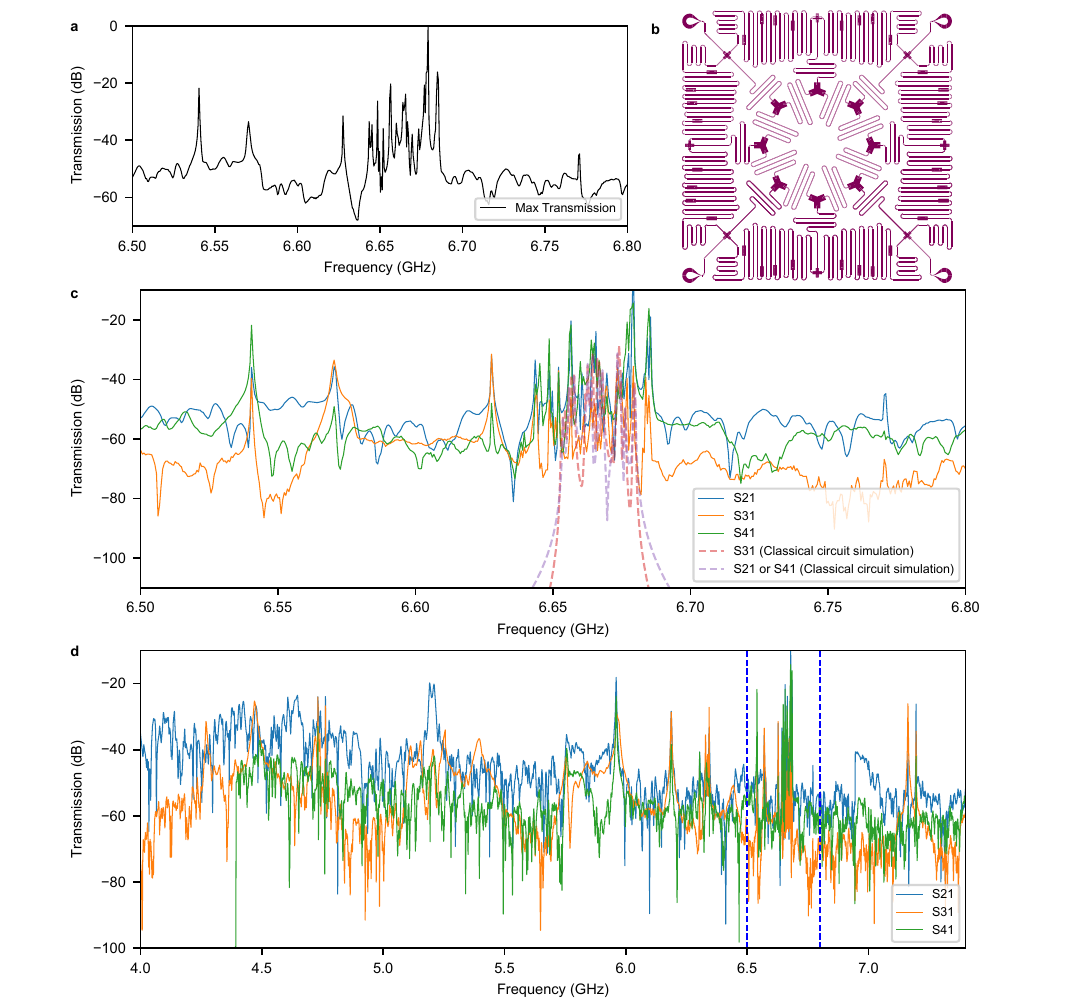}
    \caption{Data and the sample design of a kagome-like $\{8,3\}$ lattice with equal coupling between all neighbouring sites, which we call Euclidean coupling. \textbf{a}, Transmission spectra obtained by taking, at each frequency, the maximum among the measured S-parameters for $S_{21}$, $S_{31}$, and $S_{41}$. \textbf{b}, The sample design. \textbf{c}, The unprocessed S-parameters, and results from the classical circuit simulations with SPICE. Resonators are modelled as LC resonators coupled via capacitors. \textbf{d}, The same S-parameters in the widest spectrum window allowed by the cryogenic measurement setup, where the dotted blue lines show the region in (a). This plot illustrates that there is no more dense area of features apart from what is analyzed in (a).}
    \label{fig:figS7_8gon_Eu}
\end{figure*}

\end{document}